\documentclass[showpacs,twocolumn,aps,prl,amsfonts,amssymb]{revtex4}

\usepackage{graphicx}

\newcommand{\Hg}{\ensuremath{^{199}\mathrm{Hg}}}  
\newcommand{\ecm}{\ensuremath{e~\mathrm{cm}}}    
\newcommand{\sn}[2]{\ensuremath{#1 \times 10^{#2}}} 

\begin{document}

\title{Improved limit on the permanent  electric dipole moment of
$^{199}$Hg}

\author{W.~C.~Griffith}
\altaffiliation{Present address: NIST, Boulder, CO 80305}

\author{M.~D.~Swallows}
\altaffiliation{Present address: JILA, Univ. of CO, Boulder, CO 80309}

\author{T.~H.~Loftus}

\author{M.~V.~Romalis}
\altaffiliation{Permanent address: Department of Physics, Princeton
University, Princeton, NJ 08544}

\author{B.~R.~Heckel}

\author{E.~N.~Fortson}
\email[Email: ]{fortson@phys.washington.edu} \affiliation{Department of
Physics, University of Washington, Seattle, WA 98195}

\date{\today}

\begin{abstract}
We report the results of a new experimental search for a permanent electric
dipole moment of $^{199}$Hg utilizing a stack of four vapor cells.
We find $d(^{199}\mbox{Hg}) = (0.49 \pm 1.29_{stat} \pm 0.76_{syst}) \times
10^{-29}$~{\it e}~cm, and interpret this as a new upper bound,
$|d(^{199}\mbox{Hg})|$ $<$ 3.1$\times$10$^{-29}$~{\it e}~cm (95$\%$ C.L.). This
result improves our previous $^{199}$Hg limit by a factor of 7, and can be used to
set new constraints on $CP$ violation in physics beyond the standard model.
\end{abstract}

\maketitle

The existence of a finite permanent electric dipole moment (EDM) of a particle
or atom would violate time reversal symmetry ($T$), and would also imply
violation of the combined charge conjugation and parity symmetry ($CP$)
through the $CPT$ theorem \cite{Khriplovich1997,Sandars2001,Pospelov2005}. 
EDMs are suppressed in the standard model of particle physics (SM), lying many
orders of magnitude below current experimental sensitivity.  However, it is
thought that additional sources of $CP$ violation are needed to account for
baryogenesis \cite{Trodden1999,Huber2007}, and many theories beyond the SM,
such as supersymmetry \cite{Barr1993,Olive2005}, naturally predict EDMs within
experimental reach.

Experimental searches for EDMs have so far yielded null results.  The most
precise and significant limits have been set on the EDM of the neutron
\cite{Baker2006}, the electron \cite{Regan2002}, and the \Hg\  atom
\cite{Romalis2001}, leading to tight constraints on supersymmetric extensions
of the SM \cite{Olive2005}.  Here we report the first result of a new mercury
experiment, $|d(\text{\Hg})| < \sn{3.1}{-29}$~{\it e}~cm (95$\%$ C.L.), which
improves our previous limit \cite{Romalis2001} by a factor of 7 and provides a
yet more exacting probe of possible new sources of $CP$ violation.

\Hg\ has a $^1S_0$ electronic ground state and nuclear spin 1/2.  An EDM of
the ground state atom would point along the nuclear spin axis and arise mainly
from $CP$ violation in the nucleus. We measure the nuclear Larmor frequency
$\nu$ given by
$h\nu$ = $|2 \mu B \pm 2 d E|$, where $\mu$ and $d$ are the \Hg\ magnetic and
electric dipole moments, and $B$ and $E$  are the magnitudes of
external magnetic and electric fields aligned parallel ($+$) or antiparallel
($-$) with each other.  The signature for {\it d} $\neq$ 0 is thus a shift
in Larmor frequency when $\vec E$ is reversed relative to $\vec B$.

\begin{figure}[b]
\includegraphics[]{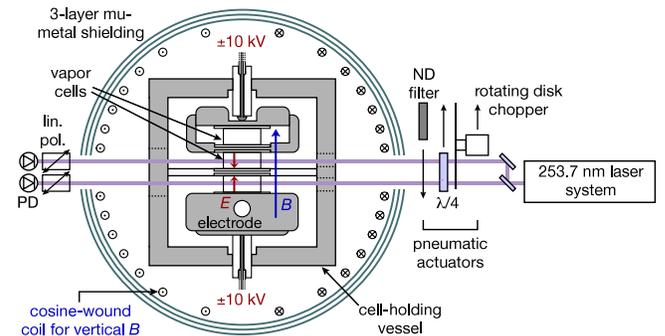}
\caption{(Color online) Simplified diagram of the $^{199}$Hg EDM apparatus
showing details of the vapor cell holding vessel and middle cell light beams.
The topmost cell is shown inside a cutaway view of the top electrode, while
the bottom electrode shows a light access hole for the enclosed cell. The
outer cell light beams are not shown but travel along the magnetic shield
axis, perpendicular to the middle cell beams. }
\label{fig:apparatus}
\end{figure}

As shown in Fig. \ref{fig:apparatus}, our new apparatus uses a stack of four
spin-polarized Hg vapor cells in a common $B$-field.  The middle two cells
have oppositely directed $E$-fields, resulting in EDM-sensitive Larmor shifts
of opposite sign; the outer two cells, enclosed by the high voltage (HV)
electrodes and thus placed at $E = 0$, are free of EDM effects and serve to
cancel $B$-field gradient noise and provide checks for spurious HV-correlated
$B$-field shifts.

The vapor cells are constructed from high purity fused silica and contain
isotopically enriched \Hg\ (92 \%) at a density of \sn{4}{13}~cm$^{-3}$, a
paraffin wall coating, and 475 Torr of CO buffer gas.  CO efficiently quenches
excited state \Hg\, and thus reduces degradation of the wall coating
\cite{Romalis2004}.  Spin coherence times $T_2$ are 100 to 200 sec. A conductive
SnO coating on the cell end-caps provides electric field plates separated by
11~mm. The average leakage currents across the cells are 0.42~pA at $\pm
10$~kV.

The vapor cells are housed inside a conductive vessel. Upper and lower
feedthroughs connect the electrodes containing the outer cells to a HV supply.
Holes in the electrodes provide optical access. All materials are free of
measurable magnetic impurities.  The vessel and electrodes are constructed
from electrically conductive, graphite-filled polyethylene. A gold coated,
fused-silica plate creates a groundplane between the two middle cells. To
minimize leakage currents, the vessel is either periodically or continuously
flushed with SF$_6$ or N$_2$.

\begin{figure}[lt]
\includegraphics[]{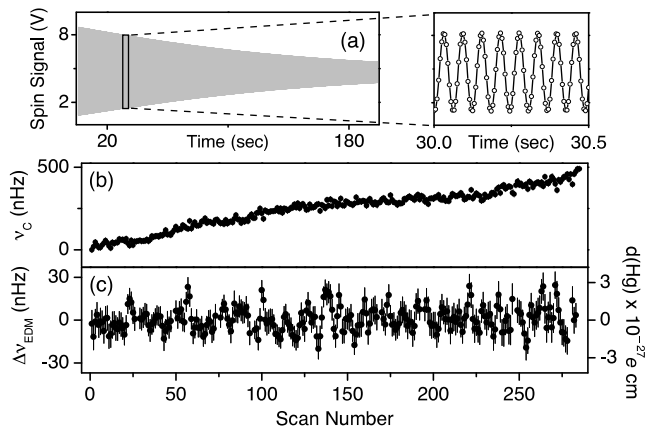}
\caption{(a) Typical single-cell precession signal with an expanded 0.5 sec
segment. (b) and (c) show $\nu_c$ and $\Delta\nu_{EDM}$ for a typical run.  In
(c) the reduced $\chi^2$ is 1.2 and the run-averaged statistical error is 0.85
nHz after scaling by $\sqrt{\chi^2}$.}
\label{DataFig1}
\end{figure}

Light for polarizing the \Hg\ spins and detecting their precession is
generated with a 254~nm laser system \cite{Harber2000} and directed through
each cell with the $\hat k$-vector perpendicular to the precession axis along
$\vec{B}$. During the 30 sec pump phase, the light is circularly polarized,
tuned to the center of the
$^{199}$Hg $^1S_0(F=1/2) \rightarrow \, ^{3}P_1 (F=1/2)$ transition, and
amplitude modulated at the 16 Hz Larmor frequency set by the 22 mG main
magnetic field, thereby building up precessing \Hg\ spin polarization
$\vec\sigma$ by synchronous optical pumping.  During the probe phase, the
light polarization is switched to linear and the frequency is tuned midway
between the $F=1/2$ and 3/2 hyperfine lines, a detuning that gives large
optical rotation angles (proportional to $\vec\sigma\cdot\hat k$), vanishing
circular dichroism, and relatively low absorption. The precession of
$\vec\sigma$ modulates the light polarization angle at the Larmor frequency;
the angle is measured, for each cell, by a photodiode after a linear
polarizer. The spin precession is monitored for 100--200 sec, after which the
pump/probe cycle is repeated. The HV is ramped to a new value during the pump
phase, typically alternating between $\pm 10$~kV.

The Larmor frequencies and their errors are extracted by fitting exponentially
decaying sine waves to the photodiode signals. Linear combinations of the
fitted frequencies are then constructed. The 
combination that maximally suppresses magnetic gradient
noise, and thus gives the best EDM sensitivity, is 
$\nu_{c} = (\nu_{MT} - \nu_{MB}) - \frac{1}{3} (\nu_{OT} - \nu_{OB})$, where
the subscripts denote cell positions:
$OT$ is outer-top, $MB$ is middle-bottom, etc.
The factor of $\frac{1}{3}$ results from the three times greater separation 
between the two outer cells compared to the middle cells.
Combinations with zero EDM sensitivity are simultaneously used to look
for systematic effects.
The EDM signal  $\Delta\nu_{EDM}$ is obtained from
 the HV-correlated component of $\nu_c$ via 3-point string analysis that
removes linear drifts in time \cite{Swallows2007}.
Figure 2 shows a single-cell precession signal
for an individual scan, along with $\nu_c$ and $\Delta\nu_{EDM}$ for a
typical run.
Data runs lasted roughly 24 hours and consisted of several hundred
individual scans.
The run-averaged statistical error for $\Delta\nu_{EDM}$ is
set by the weighted error of the mean multiplied by the square root of the
reduced $\chi^2$ (for typical
runs, $\chi^2$ was 2 or less).

Key components were periodically changed or reversed, with data taken for nine
vapor cells, four electrodes, two vessels, multiple vapor cell and electrode
orientations, and various configurations of the photodiode data acquisition
(DAQ) channels.
The vapor cell, electrode, and vessel flips used nominally identical components.
Component changes were made between groups of 10--20 runs
termed sequences; in parallel, the paraffin inside each cell was remelted and
the outer surfaces of the cells were cleaned. Each sequence comprised a
roughly equal number of dipole HV runs (+$-$+$-$ HV sequence) for the two main
$B$-field directions; one or more runs with a quadrupole HV sequence
(0+0$-$0+), sensitive to $E^2$ effects; and several tilted field (TF) dipole HV runs with the
main $B$-field tipped by $\pm10^{\circ}$ along the middle and outer light beam
$\hat k$-vectors, sensitive to $\vec{v}\times\vec{E}$ motional
$B$-fields.
A limited number of dipole HV runs were taken at 7 kV and 5 kV.  The TF runs
and two sequences at high light intensity (used to set limits on
intensity-dependent shifts) were excluded from the final value for
$\Delta\nu_{EDM}$ due to susceptibility to additional systematic errors.

An unknown, HV-correlated, EDM-mimicking offset was added to the fitted values
of $\nu_{MT}$ and $\nu_{MB}$. This fixed blind offset masked the measured EDM
and was revealed only after the data collection, data cuts, and error
analysis were complete.

\begin{figure}[rb]
\includegraphics[]{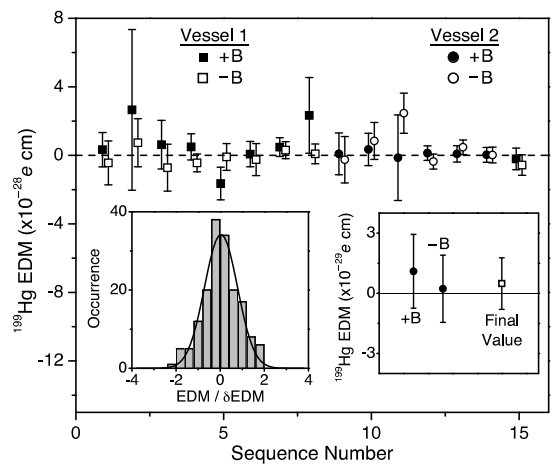}
\caption{$^{199}$Hg EDM versus sequence number. Open (closed) symbols denote
$+B$ ($-B$). Squares (circles) denote vessel 1 (vessel 2). The lower-left
inset is a histogram of $d(^{199}\mbox{Hg})$ for the 166 separate runs. The
lower-right inset shows the dataset-wide $+B$ and $-B$ values along with the
final $d(^{199}\mbox{Hg})$.}
\label{DataFig2}
\end{figure}

A potential correlation ($>$ 90$\%$ probability) was
found between $\Delta\nu_{EDM}$ and the number of micro-sparks per scan.
Nearly every spark, logged as short duration, $>$ 100~pA spikes in the
continuously monitored leakage currents, occurred in 5 sequences when the
vessel was periodically flushed with N$_2$. Two approaches to removing this
apparent correlation were 
tested: one cut individual scans where sparks occurred, the other cut entire sequences containing scans with sparks
The resulting central values for
$\Delta\nu_{EDM}$ in the two approaches 
agreed to within
1.8$\times$10$^{-30}$~\ecm. The change to
$\Delta\nu_{EDM}$ with and without spark cuts was $< 4.2 \times 10^{-30}$~\ecm.
We adopted the more conservative 
sequence-elimination approach.

Figure \ref{DataFig2} shows $d(^{199}\mbox{Hg})$ for the 15 sequences that
passed the selection criteria, along with a histogram of $d(^{199}\mbox{Hg})$
for the corresponding 166 runs. The sequence values are divided into one point
for each $B$-field direction; each point is the weighted average of the
relevant runs within the sequence. In each case, the $+B$ and $-B$ data are in
good agreement. As shown in the lower right inset, the weighted average of all
the $+B$ and $-B$ data also agree within 1-$\sigma$. Systematics that change
sign (relative to the EDM signal) when $B$ is reversed would appear in the
difference, but cancel in the average of the $+B$ and $-B$ results. Although
the data is apparently free of such problems, we determined sequence-level
values from straight averages of the $+B$ and $-B$ data. The central value for
the entire dataset was then obtained from the weighted average of the sequence
values: $d(^{199}\mbox{Hg})$ $=$ $(0.49 \pm 1.29_{stat})$$\times 10^{-29}$~{\it e}~cm.
The statistical error corresponds to a frequency difference
between the two middle cells of 0.1~nHz, a 4$\times$ improvement on Ref.
\cite{Romalis2001}.

With individual run errors set as discussed above,
the reduced $\chi^2$ for the 166 runs is 0.65. When the individual scans
are grouped into 3 hour segments and the same procedure is applied, $\chi^2
\sim 1$.
One potential source of this behavior is low frequency drift that averages
faster than white noise,
due for example to beam pointing drift that is tied to the 3 hour time scale
for resets of the piezo-actuated laser cavity mirrors.

\begin{figure}[lt]
\includegraphics[]{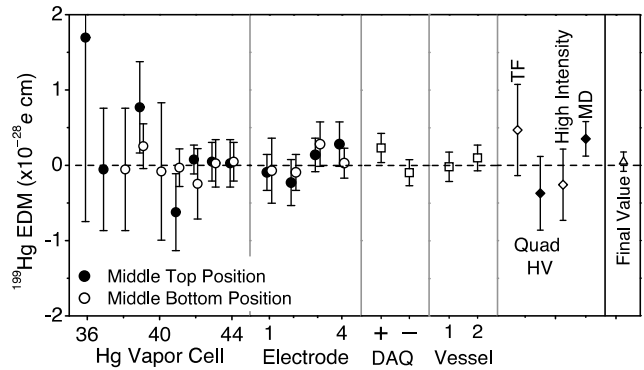}
\caption{Dependence of $d(^{199}\mbox{Hg})$ on the vapor cells, electrodes,
DAQ channel order, and vessels. 
The second pane from the right shows
$d(^{199}\mbox{Hg})$ extracted from the quadrupole HV, high intensity, and TF
runs and the two-cell difference (MD). The right-most pane shows the final
$d(^{199}\mbox{Hg})$ described in the text.}
\label{DataFig3}
\end{figure}

Figure \ref{DataFig3} summarizes several checks for systematic effects. 
We did not find statistically significant correlations between
$\Delta\nu_{EDM}$ and the vapor cells or electrodes (or their orientation
inside the vessel), the DAQ channel ordering, or the vessels.
Values for $d(^{199}\mbox{Hg})$ extracted from the TF runs, the quadrupole HV
runs, the high intensity runs, and the two-cell difference, MD = $(\nu_{MT} -
\nu_{MB})$, agree with the Fig. \ref{DataFig2} final value at the 1-$\sigma$
level.

Table \ref{Table1} summarizes the systematic errors. Three contributions
dominate. The {\it spark analysis} error is the difference in the final
$\Delta\nu_{EDM}$ value with and without spark cuts.
The {\it parameter correlations} error is obtained by multiplying the HV
correlation for key experimental parameters by the correlation of each
parameter with $\Delta\nu_{EDM}$, and then summing the products in quadrature.
Specific parameters are: the vapor cell spin amplitudes, lifetimes, relative
phases, and light transmission; the laser power, frequency, drive current, and
piezo control voltages; an external 3-axis fluxgate magnetometer; and the
$B$-field coil currents (main coil and 3 gradient coils). No statistically
significant correlations were found.
The {\it leakage current} induced error cannot be obtained from the
(unresolved) correlation slope of $\Delta\nu_{EDM}$ versus leakage current due
to the limited range over which the currents varied. A conservative estimate
is instead obtained from the worst case scenario of current flow along helical
paths within or on the vapor cell walls. The vapor cell geometry, with two
opposed seal-off stems, limits a helical current path to $\leq$ 1/2 full turn
around the cell. The average single-cell leakage current was 0.42~pA. The
resulting fields in the two middle cells can either add or subtract, so we
take $\sqrt{2}\times 0.42$~pA = 0.59~pA as the effective current. The
measurement used 9 different vapor cells (4 dominate the statistical error)
whose helical current paths should be uncorrelated; to account for this
averaging, we divide by 2. Combining the above gives a systematic error of
$4.53 \times 10^{-30}$~{\it e}~cm. The remaining Table \ref{Table1} entries
will be discussed in a longer publication.

\begin{table}[lb]
\caption[]{\label{Table1}
Systematic error budget (10$^{-30}$ {\it e} cm).}
\begin{tabular}{l@{\hspace{2mm}}c|l@{\hspace{2mm}}c}
\hline\hline Source & Error \hspace{0.3mm} & \hspace{0.3mm} Source & Error
\\\hline Leakage Currents & 4.53 \hspace{0.3mm} & \hspace{0.3mm} Charging
Currents & 0.40
\\ Parameter Correlations & 4.31 \hspace{0.3mm} & \hspace{0.3mm} Convection &
0.36\\ Spark Analysis &
4.16 \hspace{0.3mm} & \hspace{0.3mm} ($\vec{v}\times\vec{E}$) $B$-Fields &
0.18 \\ Stark Interference & 1.09 \hspace{0.3mm} & \hspace{0.3mm} Berry's
Phase & 0.18 \\ \cline{3-4} $E^2$ Effects & 0.62 \hspace{0.3mm} &
\hspace{0.3mm} {\bf Quadrature Sum} & {\bf 7.63} \\
\hline\hline
\end{tabular}
\end{table}

Summing the systematic errors in quadrature leads to our final result:
\[
d(^{199}\mbox{Hg}) = (0.49 \pm 1.29_{stat} \pm 0.76_{syst}) \times 10^{-29}
{\it e}\ {\rm cm},
\]
which we interpret as an upper limit of $|d(^{199}\mbox{Hg})|$ $<$ 
3.1$\times$10$^{-29}$~{\it e}~cm (95\% C.L.). This new limit can be
translated into upper bounds on more fundamental $CP$ violating parameters.
Table~\ref{tab:CPlimits} summarizes these limits.

\begin{table}[tr]
\caption{\label{tab:CPlimits} Limits on $CP$ violating parameters 
based on our new experimental limit for $d(\Hg)$ (95\%
C.L.) compared to limits from  the Tl (90\% C.L.) \cite{Regan2002}, neutron
(90\% C.L.) \cite{Baker2006}, or TlF (95\% C.L.) \cite{Cho1991} experiments.
Values that improve upon (complement) previous limits appear above (below) the
horizontal line.  Relevant theory references for the alternate limits
are given in the last column.
}
\begin{tabular}{lr@{$\times$}lcrr@{$\times$}lc}
\hline\hline
Parameter & \multicolumn{2}{c}{\Hg\ bound} & Hg theory & \multicolumn{4}{c}{Best
alternate limit} \\
\hline
$\tilde{d}_q$(cm)
\footnote{For \Hg: $\tilde{d}_q = (\tilde{d}_u - \tilde{d}_d)$, while for n:
$\tilde{d}_q = (0.5  \tilde{d}_u + \tilde{d}_d)$.}
& $6\;$ & $10^{-27}$  & \cite{Pospelov2002} & n: & $3 \;$ & $10^{-26}$ &
\cite{Pospelov2005} \\
$d_p$(\ecm) & $7.9\;$ & $10^{-25}$  & \cite{Dmitriev2003} & TlF: & $6 \;$ &
$10^{-23}$ & \cite{Petrov2002} \\
$C_S$ & $5.2 \;$ & $10^{-8}$  & \cite{Ginges2004} & Tl: & $2.4 \;$ & $10^{-7}$
& \cite{Sahoo2008} \\
$C_P$ & $5.1 \;$ & $10^{-7}$  & \cite{Ginges2004} & TlF: & $3 \;$ & $10^{-4}$
& \cite{Khriplovich1997} \\
\vspace{1mm}$C_T$ & $1.5 \;$ & $10^{-9}$  & \cite{Ginges2004} & TlF: & $4.5
\;$ & $10^{-7}$ & \cite{Khriplovich1997}  \\
\hline
$\bar{\theta}_{QCD}$ & $3 \;$ & $10^{-10}$ & \cite{Crewther1979} & n: & $1 \;$
& $10^{-10}$ & \cite{Pospelov2005} \\
$d_n$(\ecm) & $5.8 \;$ & $10^{-26}$ & \cite{Dmitriev2003} & \textrm{n:} & $2.9
\;$ & $10^{-26}$ & \cite{Pospelov2005} \\
$d_e$(\ecm)& $3 \;$ & $10^{-27}$  & \cite{Flambaum1985a,Martensson1987} &
\textrm{Tl:} & $1.6 \;$ & $10^{-27}$ & \cite{Ginges2004} \\
\hline\hline
\end{tabular}
\end{table}

$CP$ violation in atomic nuclei is conventionally parameterized by the Schiff
moment $S$, the lowest order $CP$ violating nuclear moment unscreened by the
electron cloud.
Atomic calculations give $d(\Hg)$ = 
$\sn{-2.8}{-17}(S/e\,\textrm{fm}^3)$~\ecm\ \cite{Dzuba2002}.
The dominant contribution to the \Hg\ Schiff moment  is expected to come from
$CP$ violating nucleon-nucleon interactions.  The most recent calculation
gives
$S(\Hg) = g(0.01\, \bar{g}^{(0)} + 0.07\, \bar{g}^{(1)} + 0.02\, \bar{g}^{(2)}) 
\,e\, \mathrm{fm}^3$  \cite{deJesus2005},
where $g$ is the strong pion-nucleon-nucleon ($\pi NN$) coupling constant, 
and $\bar{g}$ denotes
$CP$-odd $\pi NN$ isoscalar (0), isovector (1), and isotensor (2) couplings.
Note that recent work considers whether the form of $S$ used to interpret atomic
EDMs should be modified 
\cite{Liu2007,Senkov2008}.
The isovector coupling has been calculated in terms of chromo-EDMs of the
quarks, giving $\bar{g}^{(1)} = 2(\tilde{d}_u - \tilde{d}_d)
10^{14}$~cm$^{-1}$ \cite{Pospelov2002}.  For comparison, the neutron EDM is
dependent on a different combination of quark EDMs and chromo-EDMs,  $d_n =
1.1 e (0.5 \tilde{d}_u + \tilde{d}_d) +1.4 (d_d - 0.25 d_u)$
\cite{Pospelov2005}.  The quark EDMs and chromo-EDMs can be estimated in
supersymmetric extensions of the SM and in many cases, our result for
$d(^{199}\mbox{Hg})$ provides the most stringent constraints on $CP$ violating
phases \cite{Pospelov2005,Barr1993,Olive2005}.
The isoscalar coupling is dependent on the $CP$ violating phase in the QCD
Lagrangian, $\bar{g}^{(0)} \simeq 0.027 \, \bar{\theta}_{QCD}$
\cite{Crewther1979}.

Although expected to be smaller than the above effects, the proton and neutron
EDM contributions to $S$ were calculated in \cite{Dmitriev2003} giving $S(\Hg)
= (1.9 d_n + 0.2 d_p)$~fm$^2$. This allows us to set limits, $|d_n| <
\sn{5.8}{-26}$~\ecm\  and $|d_p| < \sn{7.9}{-25}$~\ecm, where in the case of
the proton we include a $30\%$ theoretical uncertainty as in
\cite{Dmitriev2003}.  This improves the best upper limit on $d_p$ by a factor
of 7.

The \Hg\ EDM can also have possible contributions from $CP$ violating
semileptonic interactions between the atomic electrons and nucleons, typically
parameterized in terms of the  
constants $C_S$, $C_P$, and $C_T$
\cite{Khriplovich1997,Ginges2004}.   New bounds on these parameters are shown
in Table~\ref{tab:CPlimits}.  We also present  a limit on the electron EDM
derived from the hyperfine structure coupling between the nuclear and
electronic spins, $d(\Hg) \simeq 0.012 \, d_e$ \cite{Martensson1987}.
This limit, however, is relatively uncertain since  an alternate estimate
is of opposite sign,
$d(\Hg) \simeq -0.014 \, d_e$ \cite{Khriplovich1997,Flambaum1985a}.

In summary, we have performed a new search for  the \Hg\ EDM.
We improve the
previous limit by a factor of 7, placing new bounds on hadronic and
semileptonic $CP$ violation. We are currently upgrading the apparatus and
expect further
improvements in the EDM sensitivity.

\begin{acknowledgments}
This work was supported by NSF Grant PHY-0457320 and the DOE Office of Nuclear
Science. We gratefully acknowledge contributions by Laura Kogler, Eric
Lindahl, David Meyer, Bob Morley, and Kristian W{\ae}degaard.
\end{acknowledgments}

\end{document}